\documentclass[prd,showpacs,preprintnumbers,amsmath,amssymb]{revtex4}
\usepackage{cancel}
\usepackage{color}
\usepackage{graphicx}
\allowdisplaybreaks

\begin{document}

\title{Heavy quark spin selection rules in meson-antimeson states}

\author{Yan-Rui Liu}\email{yrliu@sdu.edu.cn}
\affiliation{School of Physics, Shandong University, Jinan, Shandong 250100, China}

\begin{abstract}
In the heavy quark limit, we discuss the spin of the charm-anticharm pair $J_{c\bar{c}}$ in an S-wave meson-antimeson molecule or resonance. One finds two cases that $J_{c\bar{c}}$ cannot be 0: (a) $J=|j_2-j_4|-1$ or $J=j_2+j_4+1$ where $J$ is the total spin of the system and $j_2$ ($j_4$) is the total angular momentum of the light degree of freedom in a charmed meson (antimeson); (b) $J^C=1^+,2^-,3^+,\cdots$, if the two different mesons belong to the same doublet. The decays to spin-singlet charmonium states are suppressed when one of the two conditions is satisfied. We discuss constrained decay properties for selected systems.
\end{abstract}
\date{\today}
\maketitle

\section{Introduction}\label{sec1}

In the quark model, a meson contains dominantly a quark and an antiquark. As a result, the total angular momentum, the parity, and the charge conjugation parity, $J^{PC}$, of a meson cannot be $even^{+-}$, $odd^{-+}$, and $0^{--}$. However, if a meson contains four or more quarks, such exotic quantum numbers are allowed. For example, the $J^{PC}$ of an S-wave hadronic bound state of $D^*$ and $\bar{D}_2^*$ could be $3^{-+}$ if the attraction is strong enough \cite{ZhuYL13}. Just from this perspective, the study of hadronic molecules is already an interesting topic.

It is strange why only proton and neutron can form a two-body hadronic molecule, the deuteron. In the literature, there were lots of theoretical investigations about the existence of bound states in other hadron-hadron systems, but none of them has been established. Fortunately, the recent experimental findings in heavy quark hadron spectra give us chances to identify more molecules. The $X(3872)$ is the most exotic state, whose existence has been confirmed by various experimental collaborations \cite{Belle3872,CDF3872,D03872,BABAR3872,LHCb}. The state is very close to the $D^0\bar{D}^{*0}$ threshold and the molecule interpretation seems to be a natural one. However, no exotic quantum numbers may be used to identify it to be a molecule.

It is the observation of the charged $Z^\pm(4430)$ \cite{Z4430} that gives us a signal that states beyond the quark model assignment may exist. The following observations of charged states $Z_1^\pm(4051)$ \cite{Z1Z2}, $Z_2^\pm(4250)$ \cite{Z1Z2}, $Z_b^\pm(10610)$ \cite{Zb}, and $Z_b^\pm(10650)$ \cite{Zb} enhance the possibility. However, their existence has not been confirmed \cite{BABAR4430,BABARZ1Z2}. Recently, BESIII Collaboration announced the observation of a charged state $Z_c^\pm(3900)$ in the invariant mass of $J/\psi\pi^\pm$ \cite{BES3900}, which was confirmed by Belle \cite{Belle3900} and by an analysis of data taken with the CLEO-c detector \cite{Cleocdata}. The existence of multiquark states is confirmed. A popular interpretation for the structures of these charged mesons is in the molecular picture since they are close to some threshold of $Q\bar{q}$ meson and $\bar{Q}q'$antimeson. Here we do not consider their interpretations, but focus only on the meson-antimeson picture and discuss a resulting feature which is helpful to understand their structures.

Contrary to a charmonium state, where the dominant decay to another charmonium is protected by the heavy quark spin symmetry and spin-flip decays are suppressed, for a general meson-antimeson state, the decays into spin-singlet and spin-triplet charmonium states are both allowed. The reason is that the four-quark state contains both spin-0 $c\bar{c}$ component and spin-1 $c\bar{c}$ component. For instance, the amplitudes for $J_{c\bar{c}}=0$ and $J_{c\bar{c}}=1$ are equal in the two $Z_b$ states if they are $J=1$ hadronic molecules \cite{BondarGMMV11}. However, there are cases that the molecules dominantly decay into spin-triplet charmonium states. For example, Voloshin has noticed that if the $X(3872)$ is a $D\bar{D}^*$ molecule with $J^{PC}=1^{++}$, the spin of the charm-anticharm pair is only one, therefore the decay to spin-singlet states are suppressed \cite{Voloshin04}.

Once one heavy quark meson-antimeson molecule exists, more molecules are, in principle, allowed. It is possible that the spin of the $c\bar{c}$ pair in them can only be one in some cases, which is the heavy quark spin selection rule. We intend to discuss the conditions for this possibility. The discussion is helpful for the study of decay and production of the meson-antimeson states. Here the meson-antimeson state is not necessary to be a bound state.

In Ref. \cite{Voloshin04}, Voloshin argued the spin selection rule with the interpolating current and its Fierz rearrangement. In the present work, we want to construct spin wave functions at quark level and use the recoupling scheme to discuss this problem. It is easy to get heavy quark spin-0 and spin-1 amplitudes by calculating 9-j symbols. We constrain our discussions to the hidden charm case, but the results also apply to the hidden bottom analogy.

This paper is organized as follows. After the introduction in Sec. \ref{sec1}, we present the scheme for the discussions in Sec. \ref{sec2} and the results in Sec. \ref{sec3}. In Sec. \ref{sec4}, we consider the selection rules in the heavy quark baryon-antibaryon case. The final section is for discussions about strong decays of various systems. We also give a short summary there.

\section{Wave functions}\label{sec2}

Heavy quark spin symmetry plays an important role in studying properties of hadrons containing one heavy quark. In the heavy quark limit, a meson belongs to a degenerate doublet with the total spin $J=j_\ell\pm j_Q$, where $j_Q=\frac12$ is the heavy quark spin and $j_\ell$ is the total angular momentum of the light degrees of freedom. For the S-wave case, the doublet has $j_\ell=\frac12$ and the resulting mesons with $J^P=0^-$ and $1^-$ are degenerate. For the P-wave case, $j_\ell=\frac12$ or $\frac32$ and then two doublets exist: $(0^+,1^+)$ and $(1^+,2^+)$. In this paper, we will treat the charmed mesons (antimesons) as S-wave $c\bar{\tilde q}$ ($\bar{c}{\tilde q}$) bound states, where $\tilde{q}$ is an ``effective'' quark with spin $j_\ell$. Therefore, the quantum numbers of the ``quark'' in the doublets $(D,D^*)$, $(D_0,D_1^\prime)$, and $(D_1,D_2^*)$ are $j_\ell^P=\frac12^+$, $\frac12^-$, and $\frac32^-$, respectively.

To consider a meson-antimeson state in the heavy quark limit, we need wave functions and the C-parity transformations. At quark level, we use the notation
\begin{eqnarray}
D_J\sim [c_{j_1}\bar{\tilde q}_{\bar{j}_2}-(-1)^{J-j_1-\bar{j_2}}\bar{\tilde q}_{\bar{j}_2}c_{j_1}]
\end{eqnarray}
to denote the wave function of a meson and
\begin{eqnarray}
\bar{D}_J\sim [\bar{c}_{\bar{j}_1}{\tilde q}_{j_2}-(-1)^{J-\bar{j}_1-j_2}{\tilde q}_{j_2}\bar{c}_{\bar{j}_1}]
\end{eqnarray}
to denote that of an antimeson, where the subscripts indicate spin. We have added a ``bar'' to the spin of an antiquark in order to identify it just from the symbol $\bar{j}$ in the spin wave function. The factor $(-1)^{J-j_1-\bar{j}_2}$ is from the exchange of two fermions. By using these notations, we have assumed the C-parity transformation as
\begin{eqnarray}
\hat{C}:& D_J\leftrightarrow \bar{D}_J.
\end{eqnarray}

For a hidden charm state composed of a heavy quark meson $A$ and an antimeson $\bar{B}$, one may write down its wave function at quark level as
\begin{eqnarray}\label{WF-G}
X_J&\sim&[A\bar{B}+(-1)^{J-J_A-J_B}\bar{B}A]+C_X[(-1)^{J-J_A-J_B}B\bar{A}+\bar{A}B]\nonumber\\
&\sim&\Big\{[(c_{j_1}\bar{q}_{\bar{j}_2})-(-1)^{J_{12}-j_1-\bar{j}_2}(\bar{q}_{\bar{j}_2}c_{j_1})]_{J_{12}} [(\bar{c}_{\bar{j}_3}\tilde{q}_{j_4})-(-1)^{J_{34}-\bar{j}_3-j_4}(\tilde{q}_{j_4}\bar{c}_{\bar{j}_3})]_{J_{34}}\nonumber\\
&&+(-1)^{J-J_{12}-J_{34}}[(\bar{c}_{\bar{j}_3}\tilde{q}_{j_4})-(-1)^{J_{34}-\bar{j}_3-j_4}(\tilde{q}_{j_4}\bar{c}_{\bar{j}_3})]_{J_{34}}
[(c_{j_1}\bar{q}_{\bar{j}_2})-(-1)^{J_{12}-j_1-\bar{j}_2}(\bar{q}_{\bar{j}_2}c_{j_1})]_{J_{12}}\Big\}\nonumber\\
&&+C_X\Big\{(-1)^{J-J_{12}-J_{34}}[(c_{j_3}\bar{\tilde{q}}_{\bar{j}_4})-(-1)^{J_{34}-\bar{j}_3-j_4}(\bar{\tilde{q}}_{j_4}c_{j_3})]_{J_{34}}
[(\bar{c}_{\bar{j}_1}q_{j_2})-(-1)^{J_{12}-j_1-\bar{j}_2}(q_{j_2}\bar{c}_{\bar{j}_1})]_{J_{12}}\nonumber\\
&&+[(\bar{c}_{\bar{j}_1}q_{j_2})-(-1)^{J_{12}-j_1-\bar{j}_2}(q_{j_2}\bar{c}_{\bar{j}_1})]_{J_{12}}
[(c_{j_3}\bar{\tilde{q}}_{\bar{j}_4})-(-1)^{J_{34}-\bar{j}_3-j_4}(\bar{\tilde{q}}_{j_4}c_{j_3})]_{J_{34}}\Big\},
\end{eqnarray}
where both $q$ and $\tilde{q}$ indicate ``effective'' quarks and $C_X$ is the C-parity of the meson-antimeson state X. The factor $(-1)^{J-J_A-J_B}$ ($(-1)^{J-J_{12}-J_{34}}$) is from the exchange of two bosons.

We want to use the recoupling scheme to find the heavy quark spin $J_{c\bar{c}}=J_{13}$. It is enough to use the following wave function for our purpose
\begin{eqnarray}\label{WF}
X_J&\sim&\Big\{(c_{j_1}\bar{q}_{\bar{j}_2})_{J_{12}} (\bar{c}_{\bar{j}_3}\tilde{q}_{j_4})_{J_{34}}+C_X(-1)^{J-J_{12}-J_{34}}(c_{j_3}\bar{\tilde{q}}_{\bar{j}_4})_{J_{34}}
(\bar{c}_{\bar{j}_1}q_{j_2})_{J_{12}}\Big\}.
\end{eqnarray}
Now we may suppress the flavor content and just consider the spin wave function:
\begin{eqnarray}
\chi_J&=&\frac{1}{\sqrt2}\Big\{(j_1\bar{j}_2)_{J_{12}} (\bar{j}_3j_4)_{J_{34}}+C_X(-1)^{J-J_{12}-J_{34}}(j_3\bar{j}_4)_{J_{34}}
(\bar{j}_1j_2)_{J_{12}}\Big\}.
\end{eqnarray}

After recoupling the spin wave function, and noticing $\bar{j_i}=j_i$ and $(j_3\bar{j_1})_{J_{13}}=(j_1\bar{j_3})_{J_{13}}$ (the spin wave function of the $c\bar{c}$ pair), we get ($j_1=j_3=\frac12$)
\begin{eqnarray}\label{WF-chi}
\chi_J&=&\frac{1}{\sqrt2}\sum_{J_{13},J_{24}}(j_1\bar{j_3})_{J_{13}}[(\bar{j_2}j_4)_{J_{24}}-C_X(-1)^{j_2+j_4+J_{13}+J_{24}}(\bar{j_4}j_2)_{J_{24}}]\nonumber\\
&&\times\sqrt{(2J_{12}+1)(2J_{34}+1)(2J_{13}+1)(2J_{24}+1)}\left\{\begin{array}{ccc}\frac12&j_2&J_{12}\\\frac12&j_4&J_{34}\\J_{13}&J_{24}&J\end{array}\right\}.
\end{eqnarray}
If $A=B$, $(\bar{j_2}j_4)_{J_{24}}$ and $(\bar{j_4}j_2)_{J_{24}}$ must be equal, and an additional factor $\frac{1}{\sqrt2}$ should be multiplied, i.e. $\frac{1}{\sqrt2}\to\frac{1}{2}$. If $A\neq B$ but they belong to the same doublet, $(\bar{j_2}j_4)_{J_{24}}=(\bar{j_4}j_2)_{J_{24}}$ while $\frac{1}{\sqrt2}$ should not be replaced with $\frac{1}{2}$. With this formula, one may easily get the spin of $c\bar{c}$ in the meson-antimeson state and the corresponding amplitude.

Before applying the formula to various systems, we take a look at the spin wave function in the square bracket of Eq. (\ref{WF-chi}). It belongs to an ``effective'' quark-antiquark state $\bar{q}\tilde{q}$ with the total angular momentum $J_{24}$ and charge parity $c_q=C_X(-1)^{J_{13}}$. One understands the C-parity from the construction of the wave function, which is similar to that of $X_J$ in Eq. (\ref{WF-G}). The wave function of $\bar{q}\tilde{q}$ reads
\begin{eqnarray}\label{chi-q}
{[X_q]^{c_q}_{J_{24}}}\sim[\bar{q}_{\bar{j}_2}\tilde{q}_{j_4}-(-1)^{J_{24}-\bar{j}_2-j_4}\tilde{q}_{j_4}\bar{q}_{\bar{j}_2}]
+c_q[q_{j_2}\bar{\tilde{q}}_{\bar{j}_4}-(-1)^{J_{24}-\bar{j}_2-j_4}\bar{\tilde{q}}_{\bar{j}_4}q_{j_2}],
\end{eqnarray}
where $c_q$ is the C-parity. Suppress flavor content and compare it with Eq. (\ref{WF-chi}), one has
\begin{eqnarray}
[\chi_q]^{c_q}_{J_{24}}=\frac{1}{\sqrt2}[(\bar{j}_2j_4)_{J_{24}}-c_q(-1)^{J_{24}-\bar{j}_2-j_4}(\bar{j}_4j_2)_{J_{24}}]=\frac{1}{\sqrt2}[(\bar{j_2}j_4)_{J_{24}}-C_X(-1)^{j_2+j_4+J_{13}+J_{24}}(\bar{j_4}j_2)_{J_{24}}],
\end{eqnarray}
and thus $c_q=C_X(-1)^{J_{13}}$. When $q$ and $\tilde{q}$ are not equal ``effective'' quarks, a $\bar{q}\tilde{q}$ state with any $J^{PC}$ may be formed with appropriate $\bar{j}_2$ and $j_4$. One notes that the effective quark and antiquark are in different mesons and such a $\bar{q}\tilde{q}$ state is not a real meson. The state may become any meson(s) in the $X$-decay process once the quantum numbers $I^G(J^{PC})$ are conserved.

\section{Heavy quark spin selection rules}\label{sec3}

As a first example for the application of the formula, we consider the case $X\sim D\bar{D}^*\pm D^*\bar{D}$. Substitute $J_{12}=0$, $J_{34}=1$, $J=1$, $j_2=j_4=\frac12$, and $J_{24}=0,1$ into Eq. (\ref{WF-chi}) and note $(\bar{j}_4j_2)_{J_{24}}=(\bar{j}_2j_4)_{J_{24}}$, one gets
\begin{eqnarray}
\chi_J=\left\{\begin{array}{lll}
(j_1\bar{j}_3)_1(\bar{j}_2j_4)_1, &&(C_X=+)\\
\frac{1}{\sqrt2}(j_1\bar{j}_3)_0(\bar{j}_2j_4)_1-\frac{1}{\sqrt2}(j_1\bar{j}_3)_1(\bar{j}_2j_4)_0, &&(C_X=-)
\end{array}\right..
\end{eqnarray}
The spin selection rule $J_{c\bar{c}}=1$ for the $C=+$ state appears, which had been noticed by Voloshin in Ref. \cite{Voloshin04}. For $C=-$ case, there is no such a constraint, but the amplitude for spin-singlet part and spin-triplet part are equal. In the discussion of $Z_b(10610)$ \cite{BondarGMMV11}, the bottom analogy of this $J^{PC}=1^{+-}$ state, the observation has been made.

When one considers the $D^*\bar{D}^*$ state, $J_{12}=1$, $J_{34}=1$, $j_2=j_4=\frac12$, $(\bar{j}_4j_2)_{J_{24}}=(\bar{j}_2j_4)_{J_{24}}$, and $J_{24}=0,1$ are used. One finds that
\begin{eqnarray}
\chi_J=\left\{\begin{array}{lll}
\frac{1+C_X}{2}[\frac{\sqrt3}{2}(j_1\bar{j}_3)_0(\bar{j}_2j_4)_0-\frac12(j_1\bar{j}_3)_1(\bar{j}_2j_4)_1],&&(J=0)\\
\frac{1-C_X}{2}[\frac{1}{\sqrt2}(j_1\bar{j}_3)_0(\bar{j}_2j_4)_1+\frac{1}{\sqrt2}(j_1\bar{j}_3)_1(\bar{j}_2j_4)_0],&&(J=1)\\
\frac{1+C_X}{2}(j_1\bar{j}_3)_1(\bar{j}_2j_4)_1,&&(J=2)
\end{array}\right..
\end{eqnarray}
It is obvious that the spin wave function exists only when $C_X$=$+$, $-$, and $+$ for $J$=0, 1, and 2 respectively. This is consistent with the C-parity of the S-wave $D^*\bar{D}^*$ state: $C_X=(-1)^J$. The spin selection rule exists only for the case $J=2$.

If we apply the formula to the $D^*\bar{D}_1\pm D_1^*\bar{D}*$ system, where $J_{12}=J_{34}=1$, $J=0,1,2$, $j_2=\frac12$, $j_4=\frac32$, and $J_{24}=1,2$, one gets
\begin{eqnarray}
\chi_J=\left\{\begin{array}{lll}
(j_1\bar{j}_3)_1\frac{1}{\sqrt2}[(\bar{j}_2j_4)_1-C_X(\bar{j}_4j_2)_1], &&(J=0)\\
\frac{1}{\sqrt2}\Big(-\frac{1}{2\sqrt2}(j_1\bar{j}_3)_0[(\bar{j}_2j_4)_1+C_X(\bar{j}_4j_2)_1]\\
\quad+\frac{3}{4}(j_1\bar{j}_3)_1[(\bar{j}_2j_4)_1-C_X(\bar{j}_4j_2)_1]+\frac{\sqrt5}{4}(j_1\bar{j}_3)_1[(\bar{j}_2j_4)_2+C_X(\bar{j}_4j_2)_2]\Big), &&(J=1)\\
\frac{1}{\sqrt2}\Big(-\frac{\sqrt6}{4}(j_1\bar{j}_3)_0[(\bar{j}_2j_4)_2-C_X(\bar{j}_4j_2)_2]\\
\quad+\frac{1}{4}(j_1\bar{j}_3)_1[(\bar{j}_2j_4)_1-C_X(\bar{j}_4j_2)_1]+\frac{3}{4}(j_1\bar{j}_3)_1[(\bar{j}_2j_4)_2+C_X(\bar{j}_4j_2)_2]\Big), &&(J=2)
\end{array}\right..
\end{eqnarray}
The spin selection rule is there if the total angular momentum $J$ is 0.

We may perform similar calculations for other S-wave meson-antimeson states. In the calculation, one should note that the spin wave functions $(\bar{j}_4j_2)_{J_{24}}$ and $(\bar{j}_2j_4)_{J_{24}}$ are equal only when the two mesons belong to the same doublet. From the results, one finds cases that the heavy quark spin $J_{c\bar{c}}$ is only one. Table \ref{HQSR} presents some S-wave meson-antimeson states (up to P-wave mesons), the quantum numbers, and the selection rules.

\begin{table}[htb]
\begin{tabular}{cccccc}\hline
State & &$J^{C}$&$J_{24}$&Selection rule for $J_{c\bar{c}}\neq0$\\
$D\bar{D}^*/D_0^*\bar{D}_1^\prime$&&$1^{\pm}$&$0,1$&$J^C=1^+$\\
$D^*\bar{D}^*/D_1^\prime\bar{D}_1^\prime$&&$0^+,1^-,2^+$& $0,1$&$J=2$\\\hline
$D^*\bar{D}_1^\prime$&&$0^\pm,1^\pm,2^\pm$& $0,1$&$J=2$\\\hline
$D^*\bar{D}_1/D_1^\prime \bar{D}_1$&&$0^\pm,1^\pm,2^\pm$& $1,2$&$J=0$\\
$D^*\bar{D}^*_2/D_1^\prime\bar{D}^*_2$&&$1^\pm,2^\pm,3^\pm$& $1,2$&$J=3$\\\hline
$D_1\bar{D}^*_2$&&$1^\pm,2^\pm,3^\pm$&$0\sim3$&$J^C=1^+,2^-,3^+$\\
$D^*_2\bar{D}^*_2$&&$0^+,1^-,2^+,3^-,4^+$&$0\sim3$&$J=4$\\\hline
\end{tabular}
\caption{S-wave meson-antimeson states (up to P-wave mesons), quantum numbers, and selection rules for $J_{c\bar{c}}\neq0$.}\label{HQSR}
\end{table}

It is understood that the spin of the heavy quark pair is constrained by the total spin and the total angular momentum of the light degree of freedom. We summarize two selection rules for the $J_{c\bar{c}}$ in $X$ to be only one, which may be confirmed through the results in Table \ref{HQSR}:

(a) If $J=|j_2-j_4|-1$ or $J=j_2+j_4+1$, the spin of the heavy quark pair cannot be 0. This rule is easy to understand. If $J_{c\bar{c}}=0$, $J$ can only be $J_{24}$ but if $J_{c\bar{c}}=1$, $J$ can be $J_{24}+1$ or $J_{24}-1$. For maximum $J_{24}=j_2+j_4$ or minimum $J_{24}=|j_2-j_4|$, it is possible that $J$ cannot be reached if $J_{c\bar{c}}=0$.

(b) If the two different mesons belong to the same doublet, the spin of the heavy quark pair cannot be 0 when $J^C=1^+,2^-,3^+,\cdots$. We will explain this rule as follows.

We assume that the doublet contains mesons with spins $J_A$ and $J_A+1$. Then the light quark total angular momentum in a meson is $J_A+\frac12$ and that in the $X$ state is $J_{24}=0,1,\cdots,(2J_A+1)$. If $J_{c\bar{c}}=0$, the allowed $J^C$ are the same as those of light quark part in the $X$ state, namely, $0^+,1^-,2^+,\cdots,(2J_A+1)^-$. If $J_{c\bar{c}}=1$, the allowed $J^C$ are $0^+,1^\pm,2^\pm,\cdots,(2J_A+1)^\pm,(2J_A+2)^+$. Because the allowed total angular momentum of $X$ is $J=1,2,\cdots,2J_A+1$, it is observed that $J_{c\bar{c}}$ cannot be 0 when $J^C=1^+,2^-,\cdots,(2J_A+1)^+$, which is the second rule.

One may further understands the second rule through the form of S-wave decay to $\eta_c$: $X\to\eta_c+Y$. The condition for $J_{c\bar{c}}\neq0$ is equivalent to say: either $X$ or $Y$ has exotic $J^{PC}$. Here exotic means that the $J^{PC}$ are not allowed for a quark-antiquark meson. For example, in the process $1^{++}_X\to 0^{-+}_{c\bar{c}}+1^{-+}_Y$, the final state $Y$ has the exotic quantum numbers $1^{-+}$ while in the process $2^{+-}_X\to 0^{-+}_{c\bar{c}}+2^{--}_Y$, the meson-antimeson state $X$ has exotic quantum numbers. In both cases, $J_{c\bar{c}}$ cannot be 0. One may analyze this feature in a general sense. Since the involved charmed mesons have the same P-parity, the possible $J^{PC}$ of the $X$ state are $1^{+\pm},2^{+\pm},\cdots,(2J_A+1)^{+\pm}$, or $odd^{+\pm}$ and $even^{+\pm}$. For $odd^{++}_X\to 0^{-+}+odd^{-+}_Y$ ($even^{+-}_X\to 0^{-+}+even^{--}_Y$), only $Y$ ($X$) has exotic $J^{PC}$. For other two cases, $odd^{+-}_X\to0^{-+}+odd^{--}_Y$ and $even^{++}_X\to0^{-+}+even^{-+}_Y$, neither $X$ nor $Y$ has exotic $J^{PC}$. From the above analysis, $X$ with $J^{PC}=odd^{++}$ or $even^{+-}$ does not contain $J_{c\bar{c}}=0$ part. It is observed that $J_{c\bar{c}}\neq0$ occurs only when $X$ or $Y$ has exotic $J^{PC}$. It is not possible that both $X$ and $Y$ have exotic $J^{PC}$ because that case corresponds to the $0^{--}-0^{-+}-0^{+-}$ coupling but the allowed minimum angular momentum of the meson-antimeson state $X$ is 1.

However, one cannot use the form of S-wave decay to analyze possible selection rule in other cases. For example,
the quantum numbers of the $D\bar{D}_0$ state may be exotic $J^{PC}=0^{--}$, but $J_{c\bar{c}}$ can be 0 or 1. Another example is the $D\bar{D}_2^*$ state with normal $J^{PC}=2^{--}$. Although $Y$ state has exotic $J^{PC}$ in the decay $2^{--}_X\to 0^{-+}_{c\bar{c}}+2^{+-}_Y$, there is no selection rule here. The $D_0\bar{D}_1$ state has quantum numbers $J^{PC}=1^{++}$, but the spin of the charm quark pair can both be 0 and 1 although the decay $1^{++}_X\to0^{-+}_{c\bar{c}}+1^{-+}_Y$ involves exotic quantum numbers. The second rule exists because the wave functions $(\bar{j}_2j_4)_{J_{24}}$ and $(\bar{j}_4j_2)_{J_{24}}$ are equal and the constraint for $C$-parity appears. In general cases, no such constraints exist.

\section{Baryon-antibaryon case}\label{sec4}

In understanding the structure of new XYZ states, there are also suggestions that they are baryon-antibaryon bound states \cite{Qiao06}. We may also extend the previous analysis to heavy quark baryon-antibaryon case. One also gets selection rules.

In the baryon case, the wave functions of a baryon and its antipartile are denoted by
\begin{eqnarray}
B_J&\sim& [c_{j_1}(qq)_{j_2}+(-1)^{J-j_1-j_2}(qq)_{j_2}c_{j_1}],\nonumber\\
\bar{B}_J&\sim & [\bar{c}_{\bar{j}_1}\overline{({qq})}_{\bar{j}_2}+(-1)^{J-\bar{j}_1-\bar{j}_2}\overline{({qq})}_{\bar{j}_2}\bar{c}_{\bar{j}_1}].
\end{eqnarray}
Similar to the meson-antimeson case, one gets the spin wave function of a baryon-antibaryon system
\begin{eqnarray}\label{WF-chi-baryon}
\chi_J&=&\frac{1}{\sqrt2}\Big\{(j_1j_2)_{J_{12}} (\bar{j}_3\bar{j}_4)_{J_{34}}-C_X(-1)^{J-J_{12}-J_{34}}(j_3j_4)_{J_{34}}
(\bar{j}_1\bar{j}_2)_{J_{12}}\Big\}\nonumber\\
&=&\sum_{J_{13},J_{24}}(j_1\bar{j_3})_{J_{13}}[\chi_q]^{c_q}_{J_{24}}\sqrt{(2J_{12}+1)(2J_{34}+1)(2J_{13}+1)(2J_{24}+1)}
\left\{\begin{array}{ccc}\frac12&j_2&J_{12}\\\frac12&j_4&J_{34}\\J_{13}&J_{24}&J\end{array}\right\},
\end{eqnarray}
where
\begin{eqnarray}\label{chi-q-baryon}
{[\chi_q]}^{c_q}_{J_{24}}&=&\frac{1}{\sqrt2}[(j_2\bar{j_4})_{J_{24}}+C_X(-1)^{j_2+j_4+J_{13}+J_{24}}(j_4\bar{j_2})_{J_{24}}]
\end{eqnarray}
is the spinwave function of the $[(qq)\overline{\widetilde{(qq)}}]_{J_{24}}$ state, whose C-parity is $c_q=C_X(-1)^{J_{13}}$. When the baryons $[c(qq)]$ and $[c\widetilde{(qq)}]$ are the same, the factor $\frac{1}{\sqrt2}$ should be replaced by $\frac12$. The amplitude of different $J_{c\bar{c}}$ may be derived with this formula.

The difference from the meson case is that $(qq)$ is an effective boson, which makes the situation a little more complicated. In the former case, heavy quark mesons form doublets, but in the present case, besides doublets, there are two spin-singlets with $j_\ell=0$: $J^P=\frac12^+$ and $J^P=\frac12^-$ which correspond to the ground and P-wave excited $\Lambda_c$ baryons, respectively. When two singlets form a baryon-antibaryon state, obviously one has $J_{c\bar{c}}=J$. When one singlet and one baryon in a doublet form a baryon-antibaryon state, one finds that the first selection rule in the meson case still applies. For the general case, it is understandable that the first selection rule is still correct.

When two different baryons belong to the same doublet, one also gets the conclusion that $J_{c\bar{c}}$ cannot be 0 when $J^C=1^+,2^-,\cdots,(2J_A+1)^-$, where $J_A$ indicates the lower spin in the baryon doublet. Note $(2J_A+1)$ in the former case is odd while it is even now. The equivalent statement that either $X$ or $Y$ has exotic $J^{PC}$ in the form of S-wave decay $X\to\eta_c+Y$ is also correct although the P-parity of $X$ is opposite. For the decay of $X$ with exotic $J^{PC}=odd^{-+}$, the quantum numbers of $Y$ are normal $odd^{++}$. For the case of $J^{PC}=even^{--}$, it decays to $Y$ with exotic $J^{PC}=even^{+-}$. In these two cases, the spin of the charm quark pair cannot be 0. For the case of $J^{PC}=odd^{--}$ ($even^{-+}$), it decays to $Y$ state with non-exotic $J^{PC}=odd^{+-}$ ($even^{++}$)  and no selection rule exists.

Therefore, the same selection rules as the meson-antimeson case are applicable to the heavy quark baryon-antibaryon systems. The only exception is the case of two singlet baryons, where $J_{c\bar{c}}=J$.

\section{Discussions and summary}\label{sec5}

The heavy quark spin symmetry implies that the spin of the $c\bar{c}$ pair is conserved when the meson-antimeson state decays into charmonium mesons. One may use the selection rules to check the molecule interpretations for the XYZ states and identify quantum numbers from various decay channels. Note both isospin-1 and isospin-0 cases are possible for a meson-antimeson system and the spin selection rules are irrelevant to the isospin. So for a given $J^{PC}$, the selection rules are applicable to molecules or resonances with different isospins. Now we take a look at decays of various meson-antimeson systems. We restrict our discussions to the $c\bar{c}$ decay type. For convenience, we occasionally use $\psi$ to denote charmonium with $J^{PC}=1^{--}$, $J/\psi$ or $\psi(2S)$. The $\eta_c$ meson can also be replaced with $\eta_c(2S)$ if possible.

\subsection{$D\bar{D}^*$}

For $I^G(J^{PC})=0^+(1^{++})$, the state is related with the molecule interpretation of the $X(3872)$. Since $J_{c\bar{c}}=1$ here, no matter what the structure of $X(3872)$ is, a spin-triplet $c\bar{c}$ charmonium, a molecule, or a mixture of them, the decay into spin-singlet $c\bar{c}$ mesons is suppressed. Actually, such decays will not happen even the heavy quark spin symmetry is violated because of the phase space.

For $I^G(J^{PC})=1^-(1^{++})$, if such isovector states exist, they may decay into $J/\psi\pi\pi$, and $\chi_{c0,1,2}\pi$. However, a search for a charged partner of the $X(3872)$ gives negative results \cite{BABAR3872-charged}. Even if future experiments could observe such states, their decays to spin-singlet charmonium states are suppressed.

For $I^G(J^{PC})=1^+(1^{+-})$, the related state is the recently observed $Z_c^\pm(3900)$ in the $J/\psi\pi$ channel which is widely interpreted as a $D\bar{D}^*$ molecule \cite{Zc3900-th}. If the interpretation is correct, its decay channels should also include $\chi_{c0,1,2}\pi\pi$ and spin-singlet charmonium modes $\eta_c\rho$ and $h_c\pi$ since $J_{c\bar{c}}$ in the system can be 0 and 1 with equal amplitudes.

For $I^G(J^{PC})=0^-(1^{+-})$, the quantum numbers are the same as an excited spin-singlet charmonium $h_c$. If both of them exist around the threshold, they may decay into $\eta_c\omega$, but the molecule-like state can also decay into $J/\psi\eta$. It is helpful for us to study the state in the $J/\psi\eta$ channel.

The $D_1(2430)$ is a meson with $J^P=1^+$, which is identified as $D_1^\prime$. The same selection rule exists in the $D_0^*\bar{D}_1^\prime$ system and the decay products include those of $D\bar{D}^*$, but the central mass is shifted to around $4750$ MeV. Because of the broad widths of $D_0$ and $D_1^\prime$, the identification of a molecule or resonance may be rather difficult. We do not consider the system here.

One may similarly consider the hidden charm and hidden strange states $D_{s0}^*(2317)\bar{D}_{s1}(2460)$ and $D_s\bar{D}_s^*$. For the $D_{s0}^*(2317)\bar{D}_{s1}(2460)$ state with $J^{PC}=1^{++}$, it can decay into $\psi\phi$, $\chi_{c0,1,2}\eta$, and $\chi_{c0,1,2}\eta^\prime$ (no $h_c\phi$). For the state with $J^{PC}=1^{+-}$, it can decay into spin-triplet channels $\psi\eta$, $\psi\eta^\prime$, and $\chi_{c0,1,2}\phi$, and spin-singlet channels $\eta_c\phi$, $h_c\eta$, and $h_c\eta^\prime$. Because of the phase space, the $D_s\bar{D}_s^*$ state with $J^{PC}=1^{++}$ only decays into $\chi_{c0}\eta$ and the state with $J^{PC}=1^{+-}$ into $J/\psi\eta$, $\eta_c\phi$, and possibly $J/\psi\eta^\prime$ and $h_c\eta$. The spin-triplet channels in the case $J^{PC}=1^{+-}$ may be used to identify exotic states around the thresholds.

\subsection{$D^*\bar{D}^*$}

For $J^{PC}=0^{++}$, the amplitude for $J_{c\bar{c}}=0$ (coefficient=$\frac{\sqrt3}{2}$) is larger than that for $J_{c\bar{c}}=1$. The branching ratio for the decay into $\eta_c\eta$ ($\eta_c\pi$) is then expected to be larger than $J/\psi\omega$ ($J/\psi\rho$, $\chi_{c1}\pi$) in the isoscalar (isovector) case.

For $J^{PC}=1^{+-}$, the amplitudes for $J_{c\bar{c}}=0$ and $J_{c\bar{c}}=1$ are equal and the decay products include those of the $C=-$ $D\bar{D}^*$ state: $\eta_c\rho$, $h_c\pi$, $J/\psi\pi$ for isovector case and $\eta_c\omega,J/\psi\eta$ for isoscalar case. Here the $\chi_{c0,1,2}\pi\pi$ channels are also allowed.

For $J^{PC}=2^{++}$, the decay into $\eta_c$ and $h_c$ is suppressed because of the selection rule. The allowed decay channel(s) is (are) $J/\psi\omega$ ($J/\psi\rho$, $\chi_{c1}\pi$, and $\chi_{c2}\pi$) for the isoscalar (isovector) case.

Since spin is not always determined once a state around the $D^*\bar{D}^*$ threshold exists, the above mentioned channels can be used to discriminate its quantum numbers. The exotic state $Y(3940)$ was observed in the $J/\psi\omega$ invariant mass distribution \cite{BelleY3940,BABARY3940} and is thought as a $D^*\bar{D}^*$ molecule \cite{XLiuZ09,MolinaO09}. If the interpretation is correct and its angular momentum $J$ is 0, the decay channel $\eta_c\eta$ should also be observed. Otherwise, $J=2$ is favored.

After the observation of the $Z_c^\pm(3900)$, a charged state around the $D^*\bar{D}^*$ threshold is also expected. If experiments could observe such a state in the channel $J/\psi\pi$, the quantum numbers of the neutral state should be $I^G(J^{PC})=1^+(1^{+-})$. Then the state can also be observed in the $\eta_c\rho$, $h_c\pi$, and $\chi_{c0,1,2}\pi\pi$ channels. On the other hand, if experiments could observe such a state in the $J/\psi\rho$ mode, the $I^G(J^{PC})$ may be $1^-(0^{++})$ or $1^-(2^{++})$. If it can also be observed in the $\eta_c\pi$ ($\chi_{c2}\pi$) mode, one may determine the $J^{PC}$ to be $0^{++}$ ($2^{++}$), not $2^{++}$ ($0^{++}$).

 The similar system $D_s^*\bar{D}_s^*$ also has selection rule for $J=2$. The state $Y(4140)$ observed in the $J/\psi\phi$ mode is supposed to be an S-wave $D_s^*\bar{D}_s^*$ molecule \cite{XLiuZ09,GJDing09,MolinaO09}. If this picture is correct and its angular momentum $J$ is 2, its finding channel would be the main mode for $c\bar{c}$ decay. However, if it can also be observed in the $\eta_c\eta$ or $\eta_c\eta^\prime$ channel, the angular momentum $J=0$ is supported.

Because the selection rule in the $D_1^\prime\bar{D}_1^\prime$ is the same as $D^*\bar{D}^*$, the discussions are also applicable to the $D_1^\prime\bar{D}_1^\prime$ system, the central mass now is around $4860$ MeV. We do not consider the system here because of the broad width.

A little higher similar system is around the threshold of $D_{s1}(2460)D_{s1}(2460)$. If there is a state with $J^{PC}=0^{++}$, it may decay into spin-singlet channels $\eta_c\eta$, $\eta_c\eta^\prime$, $h_c\phi$, and spin-triplet channels $\psi\phi$, $\chi_{c1}\eta$, $\chi_{c1}\eta^\prime$. If the state has $J^{PC}=2^{++}$, the decay products include only spin-triplet channels $\psi\phi$, $\chi_{c1,2}\eta$, and $\chi_{c1,2}\eta^\prime$ (no $h_c\phi$). If the state has $J^{PC}=1^{+-}$, the decay channels include $\eta_c\phi$, $h_c\eta$, $h_c\eta^\prime$, $\psi\phi$, and $\chi_{c0,1,2}\phi$. One may use these different modes to discriminate quantum numbers of the state around the threshold.

\subsection{$D^*\bar{D}_1$}

For $I^G(J^{PC})=1^+([0,1,2]^{--})$, the related state is the exotic $Z(4430)$. This meson has been interpreted as an S-wave molecule \cite{XLiuLDZ08}. From Tab. \ref{HQSR}, one understands that its decay to spin-singlet charmonium processes is suppressed if this configuration is correct and the angular momentum is 0. Another allowed decay for $J=0$ is $\chi_{c1}\rho$. For $J=1$ and $J=2$, both spin-singlet and spin-triplet $c\bar{c}$ products are allowed. Since the finding channel is $\psi^\prime\pi$, it is helpful to search for this state in other modes, $h_c\pi$, $\eta_c\rho$, and $\chi_{c0,1,2}\rho$, to understand its nature. The $h_c\pi$ and $\chi_{c0}\rho$ modes couple with $J=1$. If they are not observed but other modes do, the $J=2$ is favored. If $\eta_c\rho$ and $\chi_{c2}\rho$ are also not observed, $J=0$ is favored. Of course, once the angular momentum is determined from angular distributions, the decays may be used to test the molecule model.

For $I^G(J^{PC})=0^-([0,1,2]^{--})$, the quantum numbers are the same as the excited $\psi$ states except $0^{--}$. From the selection rule, the exotic $0^{--}$ decays into $\psi\eta$, $\psi\eta^\prime$, and $\chi_{c1}\omega$ (no $\eta_c\omega$). It is not easy to identify this angular momentum from decay products. If the state around the threshold decays into $h_c\eta$ and $\chi_{c0}\omega$, it is not a charmonium and the angular momentum should be 1. If the state does not decay into $h_c\eta$ but into $\eta_c\omega$, it might be a molecule or resonance with $J=2$.

For $I^G(J^{PC})=1^-([0,1,2]^{-+})$, no related exotic state has been announced. If a state exists around the threshold, the $0^{-+}$ case can decay into $\psi\rho$ and $\chi_{c0}\pi$ (no $h_c\rho$), the $1^{-+}$ case can decay into $h_c\rho$, $\eta_c\pi$, $\psi\rho$, and $\chi_{c1}\pi$ while the $2^{-+}$ into $h_c\rho$, $\psi\rho$, and $\chi_{c2}\pi$. One may search for exotic states with these decay channels.

For $I^G(J^{PC})=0^+([0,1,2]^{-+})$, there is no mixing with the conventional spin-singlet $c\bar{c}$ meson if $J=0$. So if the decay $\chi_{c0}\eta$ is observed around the threshold, the state is very probably the pseudoscalar molecule or resonance. For $J=1$ and $J=2$, the decay into spin-triplet modes may be used to identify their exotic nature. The $1^{-+}$ state couples with $\chi_{c1}\eta$ while $2^{-+}$ couples with $\chi_{c2}\eta$.

One may similarly analyze the system $D_1^\prime\bar{D}_1$ ($D^*\bar{D}_1^\prime$) where the selection rule is for $J=0$ ($J=2$), which is ignored here. Now we turn to the allowed decay channels for similar hidden strange cases $D_s^*\bar{D}_{s1}(2536)$, $D_{s1}(2460)\bar{D}_{s1}(2536)$, and $D_s^*\bar{D}_{s1}(2460)$.

We first consider $J^{PC}=[0,1,2]^{--}$ $D_s^*\bar{D}_{s1}(2536)$. If the angular momentum is $J=0$, it decays into $\psi\eta$, $\psi\eta^\prime$, and $\chi_{c1}\phi$, but not $\eta_c\phi$ because of the selection rule. If $J=1$, the decay channels should have $\eta_c\phi$, $h_c\eta$, $h_c\eta^\prime$, $\psi\eta$, $\psi\eta^\prime$, and $\chi_{c0,1,2}\phi$. If $J=2$, the allowed modes include $\eta_c\phi$, $\psi\eta$, $\psi\eta^\prime$, and $\chi_{c1,2}\phi$. The next case is $J^{PC}=[0,1,2]^{-+}$ $D_s^*\bar{D}_{s1}(2536)$. If $J=0$, the channels include $\psi\phi$, $\chi_{c0}\eta$, and $\chi_{c0}\eta^\prime$, but not $h_c\phi$. If $J=1$ ($J=2$), the channels are $\eta_c\eta$, $\eta_c\eta^\prime$, $h_c\phi$, $\psi\phi$, $\chi_{c1}\eta$, $\chi_{c1}\eta^\prime$ ($h_c\phi$, $\psi\phi$, $\chi_{c2}\eta$, $\chi_{c2}\eta^\prime$).

For the $D_s^*\bar{D}_{s1}(2460)$ system, since the selection rule is for $J=2$, the allowed decay channels are slightly changed. (i) $J^{PC}=0^{--}$: $\psi\eta$, $\psi\eta^\prime$, $\chi_{c1}\phi$, and $\eta_c\phi$; (ii) $J^{PC}=1^{--}$: $\eta_c\phi$, $h_c\eta$, $h_c\eta^\prime$, $\psi\eta$, $\psi\eta^\prime$, and $\chi_{c0,1}\phi$; (iii) $J^{PC}=2^{--}$: $\psi\eta$, $\psi\eta^\prime$, and $\chi_{c1}\phi$. (iv) $J^{PC}=0^{-+}$: $\psi\phi$, $\chi_{c0}\eta$, $\chi_{c0}\eta^\prime$, and $h_c\phi$; (v) $J^{PC}=1^{-+}$: $\eta_c\eta$, $\eta_c\eta^\prime$, $h_c\phi$, $\psi\phi$, $\chi_{c1}\eta$, and $\chi_{c1}\eta^\prime$; (vi) $J^{PC}=2^{-+}$: $\psi\phi$, $\chi_{c2}\eta$, $\chi_{c2}\eta^\prime$, (no $h_c\phi$).

For states around the $D_{s1}(2460)\bar{D}_{s1}(2536)$ threshold, the decay modes include (i) $J^{PC}=0^{++}$: $\psi\phi$, $\chi_{c1}\eta$, and $\chi_{c1}\eta^\prime$ (no $\eta_c\eta$, $\eta_c\eta^\prime$, and $h_c\phi$); (ii) $J^{PC}=1^{++}$: $\psi\phi$, $\chi_{c0,1,2}\eta$, $\chi_{c0,1,2}\eta^\prime$, and $h_c\phi$; (iii) $J^{PC}=2^{++}$: $\psi\phi$, $\chi_{c1,2}\eta$, $\chi_{c1,2}\eta^\prime$, and $h_c\phi$; (iv) $J^{PC}=0^{+-}$: $\chi_{c0,1,2}\phi$ (no $h_c\eta$ and $h_c\eta^\prime$); (v) $J^{PC}=1^{+-}$: $\psi\eta$, $\psi\eta^\prime$, $\chi_{c0,1,2}\phi$, $\eta_c\phi$, $h_c\eta$ and $h_c\eta^\prime$; (vi) $J^{PC}=2^{+-}$: $\chi_{c0,1,2}\phi$, $h_c\eta$ and $h_c\eta^\prime$.

\subsection{$D^*\bar{D}_2^*$}

For $J^{PC}=(1,2,3)^{-+}$, the selection rule is for the highest angular momentum. In the isovector case, if $J=1$, the decay channels include the spin-triplet modes $\psi\rho$ and $\chi_{c1}\pi$, and the spin-singlet modes $\eta_c\pi$ and $h_c\rho$. If $J=2$, they are $\psi\rho$, $\chi_{c1}\pi$, and $h_c\rho$. If $J=3$, no spin-singlet modes are allowed and only $\psi\rho$ is detectable. If the $\eta_c\pi$ channel is observed, the angular momentum should be 1. One may further use the $\chi_{c1}\pi$ or $h_c\rho$ channel to identify $J=2$ if $J\neq1$. The $J=3$ can be identified finally from the $\psi\rho$ channel. In the isoscalar case, since an excited charmonium $\eta_c$ is possible, one can only use spin-triplet decay modes to identify a hadronic molecule or resonance and its angular momentum. If $\chi_{c1}\eta$ or $\chi_{c1}\eta^\prime$ ($\chi_{c2}\eta$ or $\chi_{c2}\eta^\prime$) is observed, the angular momentum $J=1$ ($J=2$) is favored. The $J=3$ can be identified through the $\psi\omega$ channel only after the exclusion of $J=1$ and $J=2$. Actually, we have studied the system in Ref. \cite{ZhuYL13}. One finds that the exotic $I^G(J^{PC})=0^+(3^{-+})$ is the most probable molecule or resonance state. The spin selection rule has been used to discuss the identification of the $3^{-+}$ state.

For $J^{PC}=(1,2,3)^{--}$, the isovector cases correspond to charged $\psi$-like states. For $J=1$, the spin wave function has significant $J_{c\bar{c}}=0$ component (coefficient=$\frac{\sqrt{10}}{4}$) and the branching ratio for the $\eta_c\rho$ and $h_c\pi$ channels is expected to be large. The state can also decay into $\psi\pi$ and $\chi_{c0,1,2}\rho$. For $J=2$, it can decay into $\eta_c\rho$, $\psi\pi$, and $\chi_{c1,2}\rho$. For $J=3$, main detectable channel would be $\chi_{c2}\rho$. So one may use these channels to discriminate the angular momentum. In the isoscalar case, it is difficult to distinguish whether the $J=3$ state is a $c\bar{c}$ meson or a molecule because of the selection rule. With the channels $h_c\eta$, $\eta_c\omega$, and $\chi_{c0}\omega$, one may discriminate the spin, $J=1$ or $J=2$.

One may perform similar analysis for the $D_s^*\bar{D}_{s2}^*(2573)$ molecules or resonances. Here we just present their possible decay channels. (i) $J^{PC}=1^{-+}$: $\eta_c\eta$, $\eta_c\eta^\prime$, $h_c\phi$, $\psi\phi$, $\chi_{c1}\eta$, $\chi_{c1}\eta^\prime$; (ii) $J^{PC}=2^{-+}$: $h_c\phi$, $\psi\phi$, $\chi_{c2}\eta$, $\chi_{c2}\eta^\prime$; (iii) $J^{PC}=3^{-+}$: only $\psi\phi$ is detectable; (iv) $J^{PC}=1^{--}$: $\eta_c\phi$, $h_c\eta$, $h_c\eta^\prime$, $\psi\eta$, $\psi\eta^\prime$, $\chi_{c0,1,2}\phi$; (v) $J^{PC}=2^{--}$: $\eta_c\phi$, $\psi\eta$, $\psi\eta^\prime$, $\chi_{c1,2}\phi$; (vi) $J^{PC}=3^{--}$: only $\chi_{c2}\phi$ is detectable.

The $D_1^\prime\bar{D}_2^*$ system has the same selection rule as the $D^*\bar{D}_2^*$ system. We ignore relevant discussions because of the broad width, but one may consider the $D_{s1}(2460)\bar{D}_{s2}^*(2573)$ states. It is found that the decay products are the same as the $D_s^*\bar{D}_{s2}^*(2573)$ molecules or resonances.

\subsection{$D_1\bar{D}_2^*$}

For $I^G(J^{PC})=1^-([1,2,3]^{++})$, only the $J=2$ state can decay into the spin-singlet channel $h_c\rho$ because of the selection rule for $J=1$ and $J=3$. In the spin-triplet modes, the $J=1$ state can decay into $\psi\rho$ and $\chi_{c0,1,2}\pi$, $J=2$ into $\psi\rho$ and $\chi_{c1,2}\pi$, and $J=3$ into $\chi_{c2}\pi$. Therefore the decay modes $h_c\rho$, $\psi\rho$, and $\chi_{c0}\pi$ may be used to discriminate the angular momentum of the state.

For $I^G(J^{PC})=0^+([1,2,3]^{++})$, since the quantum numbers are the same as excited $\chi_c$ mesons, it is not easy to distinguish a molecule from a $c\bar{c}$ meson for $J=1$ and $J=3$ cases, like the mysterious $X(3872)$. However, if a state may be observed in the $h_c\omega$ mode, the angular momentum $J=2$ is favored in the molecule structure.

For $I^G(J^{PC})=1^+([1,2,3]^{+-})$, the $J=2$ state can decay into $\chi_{c0,1,2}\rho$ but not spin-singlet mode $h_c\pi$. The $J=1$ state may decay into $\eta_c\rho$, $h_c\pi$, $\psi\pi$, and $\chi_{c0,1,2}\rho$ while $J=3$ into $\chi_{c1,2}\rho$. Once a state around the threshold is observed in the $\eta_c\rho$, $h_c\pi$, or $\psi\pi$ channel, $J=1$ is favored. If none of them but $\chi_{c0}\rho$ mode is observed, $J=2$ is favored. Only these two angular momenta are excluded, one may identify $J=3$. Actually, the spin wave function for the $J=3$ state has large $J_{c\bar{c}}=0$ component (coefficient=$\frac{\sqrt3}{2}$) and one expects large branching ratio to spin-singlet decay modes. However, it is difficult to measure such modes at present.

For $I^G(J^{PC})=0^-([1,2,3]^{+-})$, any decay mode including spin-triplet $c\bar{c}$ meson may indicate the exotic structure since the excited $h_c$ mesons dominantly decay into spin-singlet channels. With the property whether they have decay channels $\psi\eta$ and $\chi_{c0,1,2}\omega$, one may determine which angular momentum is favored. If $\psi\eta$ is observed, $J=1$ is favored. If $\chi_{c0}\omega$ is observed, $J=1$ and $J=2$ are favored.

Similar analysis may be applied to the $D_{s1}(2536)\bar{D}_{s2}^*(2573)$ states. One may easily discriminate the quantum numbers from the allowed decay channels: (i) $J^{PC}=1^{++}$: $\psi\phi$, $\chi_{c0,1,2}\eta$, and $\chi_{c0,1,2}\eta^\prime$ (no $h_c\phi$); (ii) $J^{PC}=2^{++}$: $\psi\phi$, $\chi_{c1,2}\eta$, $\chi_{c1,2}\eta^\prime$, and $h_c\phi$; (iii) $J^{PC}=3^{++}$: $\chi_{c2}\eta$ and $\chi_{c2}\eta^\prime$ (no $h_c\phi$); (iv) $J^{PC}=1^{+-}$: $\psi\eta$, $\psi\eta^\prime$, $\chi_{c0,1,2}\phi$, $\eta_c\phi$, $h_c\eta$, and $h_c\eta^\prime$; (v) $J^{PC}=2^{+-}$: $\chi_{c0,1,2}\phi$ (no $h_c\eta$ and $h_c\eta^\prime$); (vi) $J^{PC}=3^{+-}$: $\chi_{c1,2}\phi$.\\

It is difficult to search for the high spin state $J=4$ experimentally and the spin selection rule is not useful temporarily. We ignore discussions about the $D_2^*\bar{D}_2^*$ states here. Now we turn to the baryonium interpretations of $Z^\pm(4430)$, $Y(4360)$ \cite{BABARY4360,BelleY43604660}, $Y(4660)$ \cite{BelleY43604660}, and $Y(4260)$ \cite{Y4260} in Ref. \cite{Qiao06}. In this picture, the decay channels do not include the case of charmonium+one light meson.

If the $Z^+(4430)$ is the radial excitation of $\Lambda_c^+\bar{\Sigma}_c^0$, the angular momentum may be $J=0$ or $J=1$. According to the obtained spin selection rule, the $J=0$ state only decays into spin-triplet $c\bar{c}$ modes, while the $J=1$ state decays into both spin-triplet and spin-singlet $c\bar{c}$ modes. The feature is the same as the $D^*\bar{D}_1$ interpretation, but the $J=2$ is not excluded in the latter picture.

The $Y(4360)$ and $Y(4660)$ are both mixed states of $\Lambda_c\bar{\Lambda}_c$ and $\Sigma_c\bar{\Sigma}_c$ in the baryonium picture. Thus allowed decay channels include both spin-triplet and spin-singlet charmonium states. In a meson-antimeson picture \cite{Close4360}, the $Y(4360)$ is proposed as a $D^*\bar{D}_1$ bound state which can also decay into spin-singlet charmonium mesons. The two pictures have basically the same feature for decay. Since the finding channel for the $Y(4360)$ is $\pi\pi\psi(2S)$, it is helpful to understand its structure through other channels.

The $Y(4260)$ is assumed as the $\Lambda_c\bar{\Lambda}_c$ state with $J^{PC}=1^{--}$ in the baryonium picture. Therefore it only decays into spin-triplet charmonium channels. In the meson-antimeson picture, the $Y(4260)$ is a bound state of $D\bar{D}_1$ or $D_0^*\bar{D}^*$ \cite{Ding4260}. Since there is no spin selection rule, the latter interpretation indicates that the $Y(4260)$ can also decay into $h_c\eta$ and $\eta_c\omega$. The different features of the two picures may be used to understand the underlying structure.\\

The selection rules that $J_{c\bar{c}}\neq0$ with some conditions are obtained in the heavy quark limit. In reality, the heavy quark spin symmetry is not strict because of the finite heavy quark mass. The orbital-spin and spin-spin interactions between heavy quark and light quark do not vanish, the mixing among mesons with the same quantum numbers exists, and therefore the selection rules may be violated. For example, the $D^*$ meson has small D-wave portion with $j_4=\frac32$. One finds from the 9-j symbol in Eq. (\ref{WF-chi}) that $J_{c\bar{c}}=0$ is allowed now. So the decay of $1^{++}$ $D\bar{D}^*$ into $h_c$ meson is not strictly forbidden.

To summarize, we have investigated, in the heavy quark limit, the spin of the $c\bar{c}$ pair in an S-wave meson-antimeson state in the recoupling scheme. We find two cases that $J_{c\bar{c}}$ can only be one: (a) the total angular momentum $J$ is larger than the maximum angular momentum of the light degree of freedom or smaller than the minimum one; (b) $J^C=1^+,2^-,3^+,\cdots$ if the two mesons are different but belong to the same doublet. In the second case, one equivalently gets $J_{c\bar{c}}\neq0$ if either $X$ or $Y$ has exotic $J^{PC}$ in the form of S-wave decay: $X\to\eta_c+Y$. For those cases without selection rules, one may obtain the amplitude for $J_{c\bar{c}}$ from the re-coupled spin wave function of Eq. (\ref{WF-chi}). We discuss the resulting constraints for strong decays of various meson-antimeson states from these spin selection rules. The different channels may be used to discriminate the angular momentum of the state. These rules also apply to the heavy quark baryon-antibaryon states. The exceptional case then is for two singlet baryons where $J_{c\bar{c}}=J$.

\section*{Acknowledgments}

We thank Prof. Shi-Lin Zhu for helpful suggestions. This project was supported by the National Natural Science Foundation of China (No. 11275115), and Independent Innovation Foundation of Shandong University.

\end{document}